\documentstyle[aps,manuscript,epsf]{revtex}
\begin{document}
\def\de{\partial}
\def\half{{\scriptstyle{1\over 2}}}
\def\d{{\rm d}}
\newcommand{\ea}{{\it et al.\ }} 
\newcommand{\xip}{\xi^+}
\newcommand{\xim}{\xi^-}
\newcommand{\xipc}{{\xi^+}^*}
\newcommand{\ximc}{{\xi^-}^*}

\title{Decoherence of quantum wavepackets due to interaction with
conformal spacetime fluctuations} 
\author{W.L. Power, I.C. Percival}
\address{Department of Physics, Queen Mary and 
Westfield College, University of London, Mile End Road, London E1 4NS, 
United Kingdom}
\address{Telephone: +44 171 775 3292, Fax: +44 181 981 9465}
\address{email: i.c.percival@qmw.ac.uk}

\date{\today}
\maketitle

\begin{abstract}

One of the biggest problems
faced by those attempting to combine quantum theory and general
relativity is the experimental
inaccessibility of the unification scale. In this paper
we show how incoherent conformal waves in the gravitational field,
which may be produced by quantum mechanical zero-point fluctuations,
interact with the wavepackets of massive
particles. The result of this interaction is to produce decoherence
within the wavepackets which could be accessible in experiments at the
atomic scale.  

Using a simple model for the coherence properties of the gravitational
field we derive an equation for the evolution of the density matrix
of such a wavepacket. 
Following the primary state diffusion programme, the most promising
source of spacetime fluctuations for detection are the above zero-point
energy fluctuations. According to our model, the absence of
intrinsic irremoveable decoherence in matter interferometry
experiments puts bounds on some of the parameters of quantum gravity
theories. Current experiments give \( \lambda > 18. \), where \(
\lambda t_{Planck} \) is an effective cut-off for the validity of
low-energy quantum gravity theories.

\end{abstract}

\section{Introduction}

The relationship between gravity and quantum mechanics has been
studied from many different perspectives. One way of approaching the
problem is taken by those who are attempting to find a 'theory of
everything' in which gravity is a necessary component of a theory
which describes all of the fundamental forces in a unified way. This
is the point of view taken by superstring theorists
(see, for example, Green \ea 1987). A different approach to this
problem has come from those who suggest that gravity
may play a role in the quantum measurement process
(Karolyhazy 1966; Di\'{o}si 1987; Ghirardi \ea 1990; Percival 1994;
Penrose 1996),
and who use this as their starting point for looking at the quantum
properties of gravity. 

In this paper we consider conformal gravitational waves and demonstrate that
interactions with these waves can cause decoherence of quantum
mechanical wavepackets. The mechanism by which this decoherence occurs
is a consequence of the nonlinear nature of these waves. In a linear
approximation the phase shifts produced by these waves cancel.
Consequently there is no observable effect (although
observable effects have been predicted in the linear approximation if
the spacetime metric is non-commutative
(Percival \& Strunz 1997, Power 1998)). A related model
for decoherence, with non-propagating conformal fluctuations has been
studied by S\'{a}nchez-G\'{o}mez (1993). 

It has been found in earlier work on primary state diffusion that
modern matter interferometry experiments provide some of the best
tests for decoherence due to spacetime fluctuations
(Percival 1997). With this in mind we consider the possible sources
of incoherent waves in the conformal gravitational field and examine
whether these are likely to be sufficient to produce an experimentally
observable effect. One possibility that we
consider is that there may be cosmological sources of these waves, but
an analysis, based on arguments originally proposed by Rosales and
S\'{a}nchez-G\'{o}mez (1995), suggests that these are unlikely to be
observable. Another possible source of conformal 
gravitational waves are the quantum zero-point energy fluctuations in
the conformal field. By analysing these fluctuations we show that
their decoherence effect on wavepackets may be detectable in atom
interferometry experiments, although it should be noted
that by using a classical model for quantum fluctuations we cannot be
certain of including all of the essential physics. If such a
decoherence effect is not observed this may then be used to set
boundaries on some parameters of quantum gravity
theories. Improvements in the sensitivity and time-of-flight in atom
optics experiments could then be used to tighten the constraints on
these parameters. 

The structure of the paper is as follows. In section
\ref{confspacefluct} we discuss the properties of 
conformal gravitational fluctuations and describe the effects of these
fluctuations on a massive particle. In section \ref{modelflucts} we
describe the specific model which we use to describe the interaction
of a wavepacket with the conformal fluctuations and define the
coherence properties of the conformal field. In section \ref{interact}
the dynamics of the interaction are calculated, leading to an equation
for the time evolution of the density matrix of the wavepacket which
is dependent on the coherence properties of the conformal field. In
section \ref{interpret} we consider possible sources of fluctuations
in the conformal field and the amplitudes and coherence properties of
the resulting fluctuations. The equations derived in the paper are
then applied to an atom
interferometer, and it is shown that this type of experiment can be
used to put limits on the properties of the conformal fluctuations
which can then be used to put bounds on parameters appearing in some
quantum theories of gravity. Finally in section \ref{conclusions}  we
present our conclusions.

\section{Conformal spacetime fluctuations}
\label{confspacefluct}

When we describe a fluctuation as conformal what we mean is that the
deviation of the metric from the Minkowski metric is equal in all
dimensions:
\begin{eqnarray}
\label{defconf}
d \tau ^ 2 = -g_{\mu\nu} dx^{\mu} dx^{\nu} \\
g_{\mu\nu} = f(x) \eta_{\mu\nu}.
\end{eqnarray}
We choose to study conformal metric fluctuations because this simple
structure makes them easy to deal with mathematically. 

The equation for the action (in a given volume of D-dimensional spacetime) is
\begin{equation}
I(D) = -K_D\int \d^D x \sqrt{g(x)} R(x)
\label{actioninD}
\end{equation}
in which \( K_D \) is a constant depending on the number of
dimensions (for example, when \( D=4 \) then \( K_D =  {1 \over 16\pi
G} \) ), and \( R \) is the Ricci curvature scalar \cite{Kenyon1990}.

If we substitute the definition of the conformal fluctuations
(Equation \ref{defconf}) into the action integral (Equation
\ref{actioninD}) we find after some algebra that
\begin{equation}
I(D) = -{4 K_D(D-1)\over D-2}\eta^{\kappa\mu}\int\d^4 x.
          \de_\kappa A(x).\de_\mu A(x),
\end{equation}
where
\begin{equation}
A(x) = f(x)^{(D-2)/4}-1
\end{equation}
or equivalently
\begin{equation}
f(x) = (1+A(x))^{4/(D-2)}.
\label{relate}
\end{equation}
\( A(x) \) is called the conformal field amplitude. From the
variational principle we see that this field satisfies the Klein-Gordon
equation
\begin{equation}
\eta^{\kappa\mu}\de_{\kappa}\de_{\mu}A(x)=0
\end{equation}
for massless particles. When the conformal amplitude is zero 
\( g_{\mu\nu} \) 
is the Minkowski metric. Since the \( A \) field satisfies the massless
Klein-Gordon equation it describes gravitational waves
propagating at the speed of light and obeying the superposition principle.

At this point we should note that when \( A(x) = -1 \) the metric becomes
singular because \( f(x) = 0 \). For \( A(x) < -1 \) the properties of 
\( f(x) \) depend on the value of \( D \). For \( D=3,4 \), \( f(x) \) 
defines a physical metric with a positive metric tensor. For \( D =6
\), it changes the sign of the metric, and for \( D=5 \) and \( D>6 \) 
the metric is complex and non-physical. For the important case of \(
D=4 \) a real plane sinusoidal wave of amplitude greater than
unity has a set of singular planes with zero metric, but is otherwise
well-behaved. In this paper we will deal principally with waves which
have amplitudes much less than unity, and consequently these
singularities will not play an important role.

A related point is that the absolute
value of the field \( A(x) \) is important. This is unlike a potential where
we can add a constant without changing the physics.

\subsection{Massive particle in a conformal field}
\label{mconf}
Now we wish to consider the effect of the conformal field on a
particle of mass \( M \).
First suppose the particle interacts with a
general gravitational field described by a metric \( g_{\mu \nu} \). In the
Newtonian approximation and with units such that \( c=1 \), the
Lagrangian of a particle of mass \( M\) is (Appendix A) 
\begin{equation}
L_M \approx \frac{1}{2} M \dot{x}^2 - {1 \over 2} M (g_{00}(x) - 1).
\label{lagr}
\end{equation}

Next we consider only conformal spacetime fluctuations, in which case 
\begin{equation}
g_{\mu\nu}(x) = f(x) \eta_{\mu\nu},
\end{equation}
so that in the Newtonian approximation,
\begin{equation}
L_M \approx \frac{1}{2} M \dot{x}^2 - {1 \over 2} M (f(x) - 1).
\label{conflagr}
\end{equation}
In terms of the conformal amplitude (Equation \ref{relate}) this is,
\begin{equation}
L_M \approx \frac{1}{2} M \dot{x}^2 - {1 \over 2} M ((1+A(x))^{4/(D-2)} - 1).
\label{conflagr2}
\end{equation}
the action on a particle moving between the spacetime points \( x_a \) and 
\( x_b \) is then
\begin{equation}
\label{action}
I_M = \int_{x_a}^{x_b} ds \left[ \frac{1}{2} M \dot{x}^2 - {1 \over 2}
M ((1+A(x))^{4/(D-2)} - 1) \right] .
\end{equation}

\section{Model of the conformal fluctuations}
\label{modelflucts}

We now restrict our analysis to the case of a four dimensional
spacetime \( D=4 \), and consider the effects of the conformal waves
on wavepackets which are distributed along one spatial dimension. 
Our reason for doing this is that we ultimately intend to apply our model
to the special case of an atomic wavepacket separated into two
components in the arms of an interferometer and choose the spatial
coordinates so that the $x$-axis
is in the direction of the line joining the two components. However the
theory as it is derived here is more general and applies to any
one-dimensional distribution described by a density matrix \( \rho \).
In this approximation we consider that there are two types of conformal
waves, those traveling from left to right, and those propagating from
right to left. These are referred to as \( A^+(x,t) \) and \( A^-(x,t)
\) respectively. 
We assume that the intensity of the waves is such that \(
|A^{\pm}(x,t) | << 1 \), and that there is a high-frequency cut-off at
\( f_{cut-off} << 1/t_{Planck} \). This is consistent with the sources
of conformal waves considered in section \ref{interpret},
and allows us to make use of the Newtonian approximation in section
\ref{mconf}. 
We will later find that it is important to define the correlation
properties of the conformal field. We have done this by
analogy to the treatment of correlations in quantum optics
(alternative models for the correlation properties of the
gravitational field have been presented by Diosi (1988),
S\'{a}nchez-G\'{o}mez (1993) and Rosales and S\'{a}nchez-G\'{o}mez
(1995)). Separating the fluctuations into an amplitude and a fluctuating part 
\begin{eqnarray} 
A^+(x,t) \equiv A_0 \xi^+(x,t) = A_0 \xi^+_0(t-x) \\
A^-(x,t) \equiv A_0 \xi^-(x,t) = A_0 \xi^-_0(t+x),
\end{eqnarray} 
the correlation properties of the fluctuating parts are 
\begin{equation}
{\bf M} \xi^{\pm}_0(t) = 0
\end{equation}
\begin{eqnarray}
{\bf M} \xi^{s_1}_0(t) \xi^{s_2}_0(t')& =& \delta_{s_1,s_2}
g^{(1)}(t-t')\\
&=&\delta_{s_1,s_2}
{\rm exp}(-(t-t')^2/ \tau^2)
\end{eqnarray}
where \( s_i = + {\rm or} - \). \( g^{(1)} \) is the first order
correlation function, which is taken to be \( g^{(1)}(t-t')={\rm
exp}(-(t-t')^2/ \tau^2) \). The numerical value of the correlation
time \( \tau \) for physical sources of conformal fluctuations is
discussed in section \ref{interpret}. There is a further symmetry
property which is very useful:
\begin{equation}
{\bf M} \left[ \prod_{i=1..n} \xi^{\pm}_0(t^{(i)}) \right] = 0  \, \,
\mbox{ if } \, n \, \mbox{ is odd}.
\end{equation}

In addition the second order correlation function \( g^{(2)} \) is
taken as 
\begin{eqnarray}
{\bf M} (\xi^{s_1}(t))^2 (\xi^{s_2}(t'))^2& =& g^{(2)}(t-t') \; [{\rm
if} \:s_1=s_2] \\ &=& 1 \;  [{\rm if} \: s_1 \neq s_2]\\ 
&=&1+\delta_{s_1,s_2}
2{\rm exp}(-(t-t')^2/ \tau^2).
\end{eqnarray}

The choice of Gaussian correlation properties for the fluctions
simplificaties the following analysis considerably. It is
possible to work through the analysis that follows using a general
form for the fluctuations, but we choose not to do this on the grounds
of clarity.

\section{Interaction of the fluctuations with a wavepacket}
\label{interact}

The derivation of the lowest significant order in the perturbation
expansion for the interaction is {\it relatively} straightforward, but
needs some care.

The potential energy of a particle of mass \( M \) in a conformal
field \( A(x,t) \) is given by (equation \ref{action})
\begin{equation}
V(x,t)=\frac{M}{2}[(1+A(x,t))^2-1]
\end{equation}
in four-dimensional space-time. Applying the model for the propagating
fluctuations developed in the previous section gives us
\begin{eqnarray}
V(x,t)&=&\frac{M}{2} [(1 + A_0(\xip(x,t) + \xim(x,t)))^2-1] \\
&=& \frac{M}{2} A_0
[2\xip(x,t)+2\xim(x,t)+A_0(\xip(x,t)^2+\xim(x,t)^2
+2\xip(x,t)\xim(x,t))].
\label{potential}
\end{eqnarray}

Now we use perturbation theory to calculate the time evolution of a
density matrix \( \rho \) in such a potential in the time interval
from \( t=0 \) to \( t=T \). \( T \) is a long time
compared to the time taken for light to travel across the wavepacket,
and yet short compared to the time required to make a significant
change in \( \rho \).

We use a Dyson expansion to evaluate the time evolution
\begin{eqnarray}
\rho(T)&=&{\bf M}\Big(\rho(0)+K^{(1)}(T,0)\rho(0)+\rho(0)K^{(1)\dagger}
+ K^{(2)}(T,0)\rho(0) \nonumber \\
& &+ K^{(1)}(T,0)\rho(0)K^{(1)\dagger}(T,0) 
+ \rho(0)K^{(2)\dagger}(T,0) \ldots \Big)
\end{eqnarray}
and work with the terms up to-second order. The operators \( K^{(1)}
\) and \( K^{(2)} \) have the form
\begin{eqnarray}
K^{(1)}(0,T)&=&-\frac{i}{\hbar}\int_0^T H(t) dt \\
K^{(2)}(0,T)&=&-\frac{1}{\hbar^2}\int_0^T H(t)dt \int_0^t H(t') dt'. 
\end{eqnarray}

\subsection{First-order terms}

Using the form for the potential energy given by equation \ref{potential} and
neglecting (for the moment) the kinetic energy, we can start to
evaluate the terms in the Dyson expansion. We anticipate that the
first-order terms will probably lead to no overall contribution as all
parts of the wavepacket are subject to the same cumulative phase shifts
for a single conformal wave (Percival \& Strunz 1997).

The term \( {\bf M} K^{(1)}(T,0)\rho(0) \) is then given by
\begin{eqnarray}
{\bf M} K^{(1)}(T,0)\rho(0) = - \frac{i M A_0}{\hbar}
{\bf M}
\int_0^T dt \int dx
\Bigg[ 
P(x) \Big[ 2\xip(x,t)+2\xim(x,t)
+A_0\Big(\xip(x,t)^2 \nonumber \\
+\xim(x,t)^2
+2\xip(x,t)\xim(x,t)\Big) \Big]
\rho(0) \tau 
\Bigg].
\end{eqnarray}
Most terms are trivially eliminated using the property that \( {\bf M} 
\xi^{\pm}(m,n)^p = 0 \) if \( p \) is odd. Using \( \xip(x,t) =
\xip(0,t-x) \equiv \xip(t-x) \), this leaves
\begin{eqnarray}
{\bf M} K^{(1)}(T,0)\rho(0) & = & - \frac{iM A_0^2}{\hbar}
{\bf M}
\int_0^T dt \int dx
\Big[
P(x) (\xip(t-x)^2+\xim(t+x)^2) \rho(0) \tau
\Big] \\
& = & - \frac{2iM A_0^2}{\hbar} T \rho(0).
\end{eqnarray}
The evaluation of the other first-order term follows in much the same
way, and leads to the same answer except for a change of sign
\begin{equation}
{\bf M} \rho(0) K^{(1)\dagger}(T,0)
=  \frac{2iM A_0^2}{\hbar} T \rho(0).
\end{equation}

These two terms cancel, leading to no overall change in the
density matrix, as anticipated. 

\subsection{Second-order terms}

We move on to consider the second-order terms \(
K^{(2)}(T,0)\rho(0) \), \( 
K^{(1)}(T,0)\rho(0)K^{(1)\dagger}(T,0) \) and  
\( \rho(0)K^{(2)\dagger}(T,0) \). The first of these is given by
\begin{eqnarray}
{\bf M} K^{(2)}(T,0)\rho(0) = -\frac{M^2 A_0^2}{4\hbar^2} {\bf M} \Bigg(
\int_0^T dt \int dx \int_0^t dt' \int dx'
\Big[ P(x) P(x') 
\big[2\xip(x,t)+2\xim(x,t)+ \nonumber \\ A_0(\xip(x,t)^2+\xim(x,t)^2
+2\xip(x,t)\xim(x,t))\big] 
\big[2\xip(x',t')+2\xim(x',t')\nonumber\\+A_0(\xip(x',t')^2+\xim(x',t')^2
+2\xip(x',t')\xim(x',t'))\big]
\rho(0) \Big] \Bigg).
\end{eqnarray}
Many of the terms in this equation are trivially eliminated using \( {\bf M} 
\xi^{\pm}(m,n)^p = 0 \) if \( p \) is odd, leaving
\begin{eqnarray}
{\bf M} K^{(2)}(T,0)\rho(0) = -\frac{M^2 A_0^2}{4\hbar}^2 {\bf M} \Bigg(
\int_0^T dt \int dx \int_0^t dt' \int dx'
\Big[ P(x) P(x')
4\xip(t-x)\xip(t'-x') \nonumber \\
+4\xim(t+x)\xim(t'+x') 
+A_0^2 \big( 
\xip(t-x)^2\xip(t'-x')^2 
+\xip(t-x)^2\xim(t'+x')^2 \nonumber \\
+\xim(t+x)^2\xip(t'-x')^2 
+\xim(t+x)^2\xim(t'+x')^2 \nonumber \\
+4\xip(t-x)\xim(t+x)\xip(t'-x')\xim(t'+x')
\big)
\rho(0) \Big] \Bigg).
\end{eqnarray}
The projection operators simplify to give:
\begin{eqnarray}
{\bf M} K^{(2)}(T,0)\rho(0) = -\frac{M^2 A_0^2}{4 \hbar^2} {\bf M} \Bigg(
\int_0^T dt \int dx \int_0^t dt'
\Big[ P(x)
4\xip(t-x)\xip(t'-x) \nonumber \\
+4\xim(t+x)\xim(t'+x) 
+A_0^2 \big( 
\xip(t-x)^2\xip(t'-x)^2 
+\xip(t-x)^2\xim(t'+x)^2 \nonumber \\
+\xim(t+x)^2\xip(t'-x)^2 
+\xim(t+x)^2\xim(t'+x)^2 \nonumber \\
+4\xip(t-x)\xim(t+x)\xip(t'-x)\xim(t'+x)
\big)
\rho(0) \Big] \Bigg).
\end{eqnarray}

Using the expressions for the correlation functions \( {\bf M}
\xi^{s_1}(t) \xi^{s_2}(t') = 
\delta_{s_1,s_2} e^{-(t-t')^2/ \tau^2}
\) and \( {\bf M} \xi^{s_1}(t)^2
\xi^{s_2}(t')^2 = 1 + 2\delta_{s1,s2} e^{-(t-t')^2/ \tau^2}\)
as discussed in section \ref{modelflucts} gives 
\begin{eqnarray}
{\bf M} K^{(2)}_F(T,0)\rho(0) &=& -\frac{M^2 A_0^2}{4 \hbar^2} \Bigg(
\int_0^T dt \int_0^t dt' \int dx
P(x) \Big[ 
8 e^{-(t-t')^2/ \tau^2} \nonumber \\
& & +A_0^2 \big( 
4 + 4 e^{-(t-t')^2/ \tau^2} 
+4 e^{-2(t-t')^2/\tau^2}
\big) \Big]
\rho(0) \Bigg) \\
&=& -\frac{M^2 A_0^2}{4 \hbar^2} T \Big( 8 \frac{\tau \sqrt{\pi}}{2}
+ A_0^2 ( 2 T^2 + 4 T \frac{\tau \sqrt{\pi}}{2} + T \tau \sqrt{\pi} \sqrt{2} 
) \Big) \rho(0).
\end{eqnarray}
The term \( \rho(0)K^{(2)\dagger}(T,0) \) gives the same result
\begin{equation}
{\bf M} \rho(0) K^{(2)\dagger}(T,0) 
= -\frac{M^2 A_0^2}{4 \hbar^2} ( 8 \frac{T \tau \sqrt{\pi}}{2}
+ A_0^2 ( 2 T^2 + 4 T \frac{\tau \sqrt{\pi}}{2} + T \tau \sqrt{\pi} \sqrt{2} 
) ) \rho(0).
\end{equation}

Finally we come to the term \( {\bf M}
K^{(1)}(T,0)\rho(0)K^{(1)\dagger}(T,0) \). After elimination of
the simpler terms, we are left with
\begin{eqnarray}
{\bf M} K^{(1)}_F(T,0)\rho(0)K^{(1)\dagger}_F(T,0) = +\frac{M^2
A_0^2}{4\hbar}^2 {\bf M} \Bigg( 
\int_0^T dt \int dx \int_0^T dt' \int dx' 
\rho_{x,x'} \Big[
4\xip(t-x)\xip(t'-x') \nonumber \\
+4\xim(t+x)\xim(t'+x') 
+A_0^2 \big( 
\xip(t-x)^2\xip(t'-x')^2 
+\xip(t-x)^2\xim(t'+x')^2 \nonumber \\
+\xim(t+x)^2\xip(t'-x')^2 
+\xim(t+x)^2\xim(t'+x')^2 \nonumber \\
+4\xip(t-x)\xim(t+x)\xip(t'-x')\xim(t'+x')
\big)
\Big] \Bigg).
\end{eqnarray}
Using the expressions for the correlations developed earlier
this simplifies to
\begin{eqnarray}
{\bf M} K^{(1)}(T,0)\rho(0)K^{(1)\dagger}(T,0)
= \frac{M^2 A_0^2}{4\hbar^2}
\Bigg( 
\int_0^T dt \int dx \int_0^T dt' \int dx'
\rho_{x,x'} \Big[
4 e^{-(t-x-t'+x')^2/\tau^2} \nonumber \\
+4 e^{-(t+x-t'-x')^2/\tau^2}
+A_0^2 \big( 
4+2 e^{-(t-x-t'+x')^2/\tau^2}
+2 e^{-(t+x-t'-x')^2/\tau^2} \nonumber \\
+ 4e^{-(t-x-t'+x')^2/\tau^2} e^{-(t+x-t'-x')^2/\tau^2}
\big)
\Big] \Bigg).
\end{eqnarray}
In evaluating the terms in this expression we need to remember that
\( T \) is large compared to the time taken for light to traverse the
range \( r \equiv x_{max}-x_{min} \) of the wavepacket. After
evaluation of the integrals we then find
\begin{eqnarray}
{\bf M} K^{(1)}(T,0)\rho(0)K^{(1)\dagger}(T,0)
= \frac{M^2 A_0^2}{4\hbar^2} \Bigg( \Big(
8 \tau \sqrt{\pi} T 
+ A_0^2 ( 4T^2 + 4 T \tau \sqrt{\pi}) \Big) \rho(0) \nonumber \\
+ \int dx \int dx' 2 \sqrt{2} A_0^2 \tau \sqrt{\pi} T e^{-2(x-x')^2 /\tau^2}
\rho_{x,x'} \Bigg). 
\end{eqnarray}

Summing up all of the second-order terms in the perturbation expansion
leads to a significant amount of cancellation. The final result is
that to second order in perturbation theory
\begin{equation}
\label{fresult}
\rho_{x,x'}(T)=  \rho_{x,x'}(0) + \sqrt{\frac{\pi}{2}}
\frac{M^2 c^4 A_0^4 
 \tau}{\hbar^2} T
(e^{-2(x-x')^2/\tau^2}-1) \rho_{x,x'}(0), 
\end{equation}
where the dependence on \( c \) is now made explicit. 

Since the change in the density matrix is small over the time interval
\( T \) we may put this expression into a differential form 
\begin{equation}
\dot{\rho}_{x,x'}= \sqrt{\frac{\pi}{2}} \frac{M^2 c^4 A_0^4 \tau}{\hbar^2}
\Big(e^{-2(x-x')^2/\tau^2}-1 \Big)
\rho_{x,x'}
-\frac{i}{\hbar}[H_0,\rho(t)]_{x,x'}.
\label{nomorestill}
\end{equation}
where the first term represents the effect of spacetime fluctuations
and the \( H_0 \) term the rest of the Hamiltonian.

Notice that this equation is in exactly the form as appears in the
Ghiradi, Rimini and Weber (1986) theory of wavepacket reduction. 
\begin{equation}
\label{csl}
\dot{\rho}_{x,x'}= - \lambda(1-e^{-(\alpha/4)(x-x')^2})\rho_{x,x'}
-\frac{i}{\hbar}[H_0,\rho(t)]_{x,x'}.
\end{equation}
This equation first appeared 
in the work of Barchielli \ea (1982), where
it describes the evolution of a continuously observed
quantum wavepacket. In the GRW model a quantum
wavepacket undergoes occasional instantaneous quantum jumps which localise
the position of the quantum system, in which case the ensemble average
obeys this same density matrix equation. This equation also
appears in the theory of continuous spontaneous localisation where the
quantum wavepacket is continuously subject to interactions which
localise its position (Di\'{o}si 1988, Gisin 1989, Pearle 1989). 

We have identified a physical mechanism for equation \ref{csl} and
supplied values for the undetermined constants
\begin{equation}
\lambda = \sqrt{\frac{\pi}{2}} \frac{M^2 c^4 A_0^4 \tau}{\hbar^2}
\end{equation}
and
\begin{equation}
\alpha = 8/\tau^2.
\end{equation}

This result is dependent on our previous choice of Gaussian correlation
properties for the fluctuations; for the result in the general case of an
unspecified correlation function see Appendix
\ref{generalresult}. Pearle and Squires (1996) have
made a different estimate of the parameters \( \alpha \) and \(
\lambda \), starting with the theory of continuous spontaneous localisation.

\section{Interpretation}
\label{interpret}

At this stage we must consider the possible sources of
conformal fluctuations. There may be cosmological sources, the
amplitude and frequency of which are hard to 
predict. Arguments given by Rosales and S\'{a}nchez - G\'{o}mez (1995)
set a limit on
the energy density that can be present in the form of conformal waves without
adding sufficient extra mass to the universe that the universe would
collapse in on itself. According to their argument the energy density
of conformal waves must not exceed \(10^{-29} {\rm g/cm^{3}} \), and,
they argue, this limits the correlation length to being greater than
\( 10^{-3} {\rm cm} \) equivalent to having a correlation time of
greater than \( 10^{-13} {\rm s} \). Using dimensional arguments they
relate the amplitude of the conformal fluctuations to the correlation
length to set a maximum amplitude of \( 10^{-30} \). Substitution of
these values into equation \ref{fresult} suggests that the effects of
these conformal waves would then probably never be observed in a
interferometry experiment.

A second possible source of conformal waves is the quantum zero-point
fluctuations of the gravitational field. 
We assume here that the classical conformal waves correspond to the
zero-point fluctuations of quantum gravity, but also that the energy
density normally associated with such classical waves is canceled by
renormalization. Otherwise the fluctuations would be ruled out on the
basis of the arguments of Rosales and S\'{a}nchez-G\'{o}mez.

According to many theories, such as
the superstring theory, there is a length scale below which the
universe in no longer effectively four-dimensional but has a higher
number of dimensions. We may suppose that this length scale may
effectively provide a cut-off wavelength for the conformal
gravitational waves. Let us call this cut-off length \( l_{cut-off} \)
and let us relate this to the Planck length \( l_{Planck} \) and define \(
 \lambda \) such that 
\begin{equation}
\label{lambdadefinition}
l_{cut-off} = \lambda l_{Planck}.
\end{equation} 
This defines a cut-off frequency \( \omega_M = 2 \pi c
\lambda_{cut-off}^{-1} \).
The amplitude of the zero-point conformal fluctuations in the
gravitational field is then approximately (see Appendix \ref{zeropoint})
\begin{equation}
\label{fluctscale}
A_0 \approx \omega_M^2 t_{Planck}^2 \approx \lambda^{-2}.
\end{equation}

Current theories place the value of \( \lambda \) in the range of \(
10^{2} - 10^{6} \). Within this range the Newtonian approximation that
was made in section \ref{mconf} should be adequate for the
order-of-magnitude estimates that we are interested in here. 

Now we consider a matter interferometer which puts particles of mass
\(M\) into a superposition of states, separated by a distance which is very
large compared to the correlation length of the conformal
fluctuations, for a time \( T \). Then from equation
\ref{fresult} we see that  
\begin{equation}
\frac{\delta \rho}{\rho(0)}=\sqrt{\frac{\pi}{2}} \frac{M^2 c^4 A_0^4
\tau T}{\hbar^2}.
\end{equation}
By substitution of equation \ref{fluctscale} and setting the
correlation time \( \tau = \lambda t_{Planck} \) (which follows from
equation \ref{lambdadefinition} when the fluctuations travel at the
speed of light) we find that \( \lambda \) is 
\begin{equation}
\label{lambdaest}
\lambda = \Bigg( \sqrt{\frac{\pi}{2}}\frac{M^2c^4t_{Planck}T}{\hbar^2
(\delta \rho / \rho(0))} \Bigg)^{1/7}.
\end{equation} 
In practice there will be other experimental causes for loss of
contrast in the interference pattern, such as vibrational instabilities
in the apparatus. Hence, by substituting the experimentally determined
values for the loss of contrast into equation \ref{lambdaest} we
find a lower bound for \( \lambda\). Notice that due to the \( 1/7\)th
power in equation \ref{lambdaest} the bound on \( \lambda \) only
rises slowly with improvements in the experimental parameters.

In recent experiments by Peters \ea (1997) Caesium atoms
(relative atom mass 132.9) were put
into a superposition state for \(0.32 {\rm s} \) before recombining
to form an interference pattern. The observed loss
of contrast over this period was approximately \(3 \% \). Substitution
of these parameters into equation \ref{lambdaest} leads to the setting
of a lower bound on \( \lambda \) of 
\begin{equation}
\lambda > 18.
\end{equation} 
Improvements in the time and
resolution of atom interferometry experiments should enable the bound
on \( \lambda \) to be raised. It is interesting to contrast this
behaviour with that of particle accelerators which are effectively
able to reduce the upper bound on \( \lambda \) by increasing particle
energies, although current experiments still put an upper bound on
\(\lambda\) well above the range of theoretical predictions.

\section{Conclusions}
\label{conclusions}
We have derived the leading terms in an expansion for the nonlinear
interaction between the conformal gravitational field and wavepackets
of massive particles. When the field contains incoherent conformal waves we 
have shown that this leads to the
decoherence of the wavepackets. We have
considered possible sources of conformal gravitational waves, and have
shown how atom interferometry experiments can be used to try to detect their
presence. If no decoherence is observed in these
experiments then our model can be used to place limits on the
properties of sources of these waves. If we have a theory for the
sources of these waves, such as the model for quantum zero-point
fluctuations presented here, then these limits can be used to place
boundaries on the fundamental properties, such as the size of
compactified dimensions, present in those theoretical models.

The authors would like to thank J. Charap, S. Thomas, W. Strunz and
the QMW superstring group for very helpful discussions. We are also
grateful to the EPSRC and the Leverhulme Foundation for providing the
financial support that made this work possible. 

\appendix
\section{Derivation of Lagrangian}
\label{deriv}Consider the motion of a mass \( M \) in a gravitational field
specified by a metric \( g_{\mu \nu} \). The general geodesic equation
for the velocity 
vector \( v^{\mu}=dz^{\mu}/ds \) is 
\begin{equation}
{dv^{\mu}\over{ds}}+\Gamma^{\mu}_{\nu\sigma}v^{\nu}v^{\sigma} = 0.
\label{geod}
\end{equation}
If the field is slowly varying in time
such that 
\begin{equation}
 \de_0 g_{\mu\nu} \approx 0,
\label{cond1} 
\end{equation}
and the field is assumed
to be weak so that
\begin{equation}
g_{\mu\nu} = \eta_{\mu\nu} + h_{\mu\nu}
\label{cond2}
\end{equation}
where all the components of the tensor \( h\) are much less than \( 1
\). Then for a particle moving slowly compared to the velocity of
light
\begin{equation}
|v^m| \ll 1, \; (m = 1,2,3)
\label{cond3}
\end{equation}
we find by substitution into Equation \ref{geod} and neglecting second
order quantities that
\begin{equation}
{dv_m \over{dx^0}}= \de_m (g_{00})^{1/2}.
\end{equation}
From this it appears that the particle is moving in a potential 
\( (g_{00})^{1/2} \); if we write
\begin{equation}
g_{00}(x) = 1 + 2 U_{G}(x)
\end{equation}
then for small \( U_{G} \)
\begin{equation}
(g_{00})^{1/2} \approx 1 + U_{G}(x)
\end{equation}
and we see that \( U_{G}(x) \) is the classical gravitational
potential. The gravitational potential energy of the particle of mass
\( M \) is therefore
\begin{equation}
M U_{G} \approx {1 \over 2} M (g_{00}(x) - 1).
\end{equation}
The Lagrangian of the particle, provided the above conditions hold
(Equations \ref{cond1} to \ref{cond3}), is then
\begin{equation}
L_M \approx \frac{1}{2} M \dot{x}^2 - {1 \over 2} M (g_{00}(x) - 1).
\label{lagr2}
\end{equation}

\section{Evolution of matrix elements with generalized fluctuations}
\label{generalresult}

If instead of assuming Gaussian correlation properties for the
fluctuations, we work with a general expression for the fluctuations
we find, after a considerable amount of algebra along the lines of
section \ref{interact}, that
\begin{eqnarray}
\rho_{x,x'}(T) \approx \rho_{x,x'}(0) + \frac{M^2A_0^4}{\hbar^2} \Bigg( \int_0^T dt
\int_0^T dt' g^{(1)}(t-t'-x+x') g^{(1)}(t-t'+x-x')
\rho_{x,x'}(0) \nonumber \\ - 2 \int_0^T dt \int_0^t dt' \Big(g^{(1)}(t-t')\Big)^2
\rho_{x,x'}(0) \Bigg)
\end{eqnarray}
which when the correlation function is taken to be Gaussian 
\begin{eqnarray}
{\bf M} \xi^{s_1}_0(t) \xi^{s_2}_0(t')& =& \delta_{s_1,s_2}
g^{(1)}(t-t')\\
&=&\delta_{s_1,s_2}
{\rm exp}(-(t-t')^2/ \tau^2)
\end{eqnarray}
reduces to equation \ref{fresult}:
\begin{equation}
\rho_{x,x'}(T)=  \rho_{x,x'}(0) + \sqrt{\frac{\pi}{2}} \frac{M^2 A_0^4
\tau}{\hbar^2} T
(e^{-2(x-x')^2/\tau^2}-1) \rho_{x,x'}(0). 
\end{equation}

Note that the precise form of the second order correlation function is
unimportant; all that we require is that \( g^{(2)}(t) \) should be an
even function and that \(
\int_{-\infty}^{\infty} (g^{(2)}(t)-1) dt \) should be finite. Both of these
conditions are features of physically realistic second-order
correlation functions.

\section{Evaluation of the amplitude of zero-point conformal
gravitational waves}
\label{zeropoint}

We first summarise the properties of conformal gravitational waves as
derived in Section \ref{confspacefluct}
\begin{eqnarray}
d \tau ^ 2 = -g_{\mu\nu} dx^{\mu} dx^{\nu} \\
g_{\mu\nu} = f(x) \eta_{\mu\nu}.
\end{eqnarray}
and
\begin{equation}
f(x) = (1+A(x))^{4/(D-2)}
\end{equation}
where \( A(x) \) is called the conformal field amplitude. This
satisfies the Klein-Gordon equation
\begin{equation}
\label{apkg}
\eta^{\kappa\mu}\de_{\kappa}\de_{\mu}A(x)=0
\end{equation}
for massless particles. When the conformal amplitude is zero 
\( g_{\mu\nu} \) 
is the Minkowski metric.
The Klein-Gordon equation is also known as the wave equation, and
equation \ref{apkg} can be written as
\begin{equation}
\label{apwav}
\frac{1}{c^2}\frac{\partial^2 A(x,t)}{\partial t^2}=\nabla^2 A(x,t).
\end{equation}

This is essentially the same as the equation for the components of the
electromagnetic potential in flat space, and the same methods can be
used to obtain the energy density (see, for instance Loudon 1983).

By quantizing in a cube of volume \( V \), we find that for the scalar
field \( A \) the density of modes (not taking into account
supersymmetry) with frequency between
\( \omega \) and 
\( \omega + d\omega \) is 
\begin{equation}
4\pi \frac{\omega^2 d\omega}{(2 \pi c)^3}V.
\end{equation}
If each mode has a zero point energy of \( \frac{1}{2}\hbar\omega \) and we
have a cut-off frequency of \( \omega_M \), then the total energy per unit
volume is 
\begin{equation}
\int_0^{\omega_M} \frac{1}{2} \hbar \omega 4\pi \frac{\omega^2 d\omega}{(2 \pi
  c)^3} d \omega = \frac{1}{16} \frac{\hbar \omega_M^4}{\pi^2 c^3}
\end{equation}
Assuming that the squared
amplitude of the gravitational field fluctuations is proportional to the
energy density tells us that \( |A_0| \propto \omega_M^2 \). As we
expect the (dimensionless) amplitude to be of the order of unity if the
cut-off is set to be the Planck length this leads us to deduce that
\begin{equation}
|A_0| \approx \omega_M^2 t_{Planck}^2.
\end{equation}

\end{document}